\documentclass[pra,twocolumn,aps,superscriptaddress,showpacs,floatfix]{revtex4-1}
\usepackage{graphicx,amsfonts,amssymb,amsmath,hyperref,hypcap,enumerate}
\usepackage{epsfig}
\usepackage{subfigure}
\usepackage{mathtools}
\usepackage{braket}
\usepackage[usenames,dvipsnames]{color}
\usepackage{setspace}
\usepackage{bm}
\usepackage{ulem}
\usepackage{times}
\usepackage{enumitem}
\usepackage{xcolor}
\usepackage{multirow}
\usepackage{soul}
\usepackage{dcolumn}   
\usepackage{lineno}
\usepackage{balance}
\usepackage{epstopdf}
\usepackage{color}
\usepackage{diagbox}
\usepackage[utf8]{inputenc}
\hypersetup{
      colorlinks=true,
      citecolor=blue,
      linkcolor=blue,
      urlcolor=blue}

\DeclareMathAlphabet{\mathitb}{OT1}{cmr}{bx}{sl}

\def\Tr{\mathrm{Tr}}
\def\be{\begin{equation}}
\def\ee{\end{equation}}
\def\bg{\begin{equation}\begin{gathered}}
\def\eg{\end{gathered}\end{equation}}

\begin{document}
%\pagewiselinenumbers
\title{Non-Hermitian Many-Body Localization with Open Boundaries}
\author{Kuldeep Suthar}
\email[Corresponding author. ]{kuldeep@gate.sinica.edu.tw}
\affiliation{Institute of Atomic and Molecular Sciences,
             Academia Sinica, Taipei 10617, Taiwan}
\author{Yi-Cheng Wang}
\affiliation{Institute of Atomic and Molecular Sciences,
             Academia Sinica, Taipei 10617, Taiwan}
\affiliation{Department of Physics, National Taiwan
             University, Taipei 10617, Taiwan}
\author{Yi-Ping Huang}
\affiliation{Department of Physics, National Tsing Hua
             University, Hsinchu 300044, Taiwan}
\affiliation{Institute of Physics,
             Academia Sinica, Taipei 115, Taiwan}
\author{H. H. Jen}
\email[Corresponding author. ]{sappyjen@gmail.com}
\affiliation{Institute of Atomic and Molecular Sciences,
             Academia Sinica, Taipei 10617, Taiwan}
\affiliation{Physics Division, National Center for Theoretical Sciences, 
             Taipei 10617, Taiwan}
\author{Jhih-Shih You}
\email[Corresponding author. ]{jhihshihyou@ntnu.edu.tw}
\affiliation{Department of Physics, National Taiwan Normal
             University, Taipei 11677, Taiwan}

\date{\today}

%%%%%%%%%%%%%%%%%%%%%%%%%%%%%%%%%%%%%%%%%%%%%%%%%%%%%%%%%%%%%%%%%%%%%%%%%%%%%%
%%%%%%                             Abstract                               %%%%
%%%%%%%%%%%%%%%%%%%%%%%%%%%%%%%%%%%%%%%%%%%%%%%%%%%%%%%%%%%%%%%%%%%%%%%%%%%%%%
\begin{abstract}
The explorations of non-Hermiticity have been devoted to investigate the 
disorder-induced many-body localization~(MBL). However, the sensitivity of the 
spatial boundary conditions and the interplay of the non-Hermitian skin effect 
with many-body phenomena are not yet clear. For a MBL system in the presence of 
non-reciprocal tunnelings and random disorder potential, we identify two 
different complex-real spectral transitions, one is present for both open and 
periodic boundaries while the other is present only for open boundaries of a 
coupled non-Hermitian chains. The later is driven due to the inter-chain 
coupling at weak disorder where the level statistics of the real eigenenergy 
phase follows Gaussian orthogonal ensemble. We further characterize 
wavefunctions through the~(biorthogonal) inverse participation ratio and fractal 
dimension, which reveal the suppression of skin effect in the non-Hermitian MBL 
phase. Finally, we demonstrate that the quench dynamics of the local particle 
density, spin imbalance, and entanglement entropy also signify the hallmark of 
the boundary effects and non-ergodic character of many-body localization.
\end{abstract}
\maketitle

%%%%%%%%%%%%%%%%%%%%%%%%%%%%%%%%%%%%%%%%%%%%%%%%%%%%%%%%%%%%%%%%%%%%%%%%%%%%%%
%%%%                             Introduction                          %%%%%%%
%%%%%%%%%%%%%%%%%%%%%%%%%%%%%%%%%%%%%%%%%%%%%%%%%%%%%%%%%%%%%%%%%%%%%%%%%%%%%%
\section{Introduction}
 When traditional quantum mechanics postulates Hermiticity, numerous 
developments have been made for exploring non-Hermitian quantum mechanics in 
various fields of physics like condensed-matter, cold atoms, and open quantum 
systems~\cite{bender_98,longhi_09,yuto_17,gong_18,ashida_20,bergholtz_21}. 
The recent experimental advances provide access to engineer the non-Hermitian 
Hamiltonians with the dissipation and non-reciprocal 
tunnelings~\cite{xiao_20,helbig_20,hofmann_20,gou_20,liang_22}. These 
developments allow us to explore fundamental physics of localization. The 
interplay between disorder and non-Hermiticity due to asymmetric hopping was 
first investigated by the pioneer works of Hatano and Nelson, which reveal a 
real-complex transition of single-particle 
spectrum~\cite{hatano_96,hatano_97,hatano_98}. Moreover, it has been shown that 
the random potential can suppress the complex eigenenergies of an interacting 
single Hatano-Nelson chain with periodic boundary condition~(PBC) and the 
spectral transition is accompanied with non-Hermitian MBL 
transition~\cite{hamazaki_19}. The similar phenomena have also been found for 
quasiperiodic potential~\cite{zhai_20,Orito_22}. Since the disordered 
single-chain model respecting time-reversal symmetry belongs to the symmetry 
class AI, its localized phase follows the real Poisson ensemble while the 
delocalized phase follows the 
Ginibre ensemble~\cite{ginibre_65,grobe_88,haake_13}.

The choice of the imposed boundary conditions plays decisive role in determining 
the properties of non-Hermitian systems. One fascinating phenomenon that has no 
Hermitian counterparts is the non-Hermitian skin effect (NHSE). It describes 
the anomalous localization for an extensive number of bulk modes which can 
occur at the boundaries of non-Hermitian open 
lattices~\cite{yao_18,alvarez_18,borgnia_20,okuma_20,zhang_20,yoshida_20,
kawabata_20,ryo_20,fu_21,zhangkai_22,wang_21}. This effect fundamentally 
challenges our knowledge of the band theory and violates the conventional bulk 
boundary correspondence of Hermitian systems which connects robust edge states 
to bulk topological invariants. For non-Hermitian Hamiltonians, the 
understanding of NHSE in non-interacting systems relies on the non-Bloch theory 
in which non-Bloch topological invariants are defined in generalized Brillouin 
zones~\cite{yokomizo_19,bergholtz_21}. The NHSE has been realized in various 
experimental setups of photonics~\cite{xiao_20,weidemann_20}, electrical 
circuits~\cite{helbig_20,liu_21,hofmann_20}, 
metamaterials~\cite{brand_19,ghatak_20}, and ultracold 
atoms~\cite{gou_20,liang_22}. More recently, the fate of skin modes in fermionic 
and bosonic systems with many-body correlations has been 
investigated~\cite{lee_20,mu_20,tsuneya_21,zhang_21,liu_20,cao_21,zhang_20a,
xu_20,faisal_21,zhang_22,kawabata_22}. However, little is known about the 
robustness of NHSE on the localization properties of many-body disordered 
systems.

To date, the unusual characteristics of non-Hermitian systems have been explored 
from various perspectives. One of them is the sensitivity of boundary conditions 
which leads to the emergence of NHSE. Under PBC, for a single periodic 
non-Hermitian chain with non-reciprocal tunnelings, the many-body localization 
transition occurs with a complex-real spectral 
transition~\cite{hamazaki_19,zhai_20}. However, this does not apply to a single 
chain under open boundary condition (OBC) due to the real spectrum in the 
presence of NHSE. Recently, many theoretical studies have uncovered that the 
non-trivial behaviour arises due to the finite coupling between two one-band 
subsystems with different generalized Brillouin 
zones~\cite{li_20,li_20a,yokomizo_21,mu_21,rafi_22}. It is legitimate to ask 
whether the boundary-induced phenomena exist when two subsystems are coupled 
together with unequal (or equal) non-reciprocal hopping parameters. The 
interplay of the non-reciprocal hoppings and coupling strength of two-chain for 
many-body disordered system is yet to be explored.

In this work, we study the spectral statistics, eigenstate properties, and 
non-equilibrium dynamics of many-body coupled Hatano-Nelson chains in the 
presence of a random disorder potential. In particular, we investigate the role 
of the imposed boundary conditions on the characteristic properties of a 
non-Hermitian two-chain system. We identify a parameter space of inter-chain 
coupling and non-reciprocal tunneling parameters, where a complex-real spectral 
transition is observed for finite open chains, while the spectrum for PBC 
maintains complex. The spectral transition is accompanied by the change in 
nearest-level spacing distribution of eigenenergies and average complex spacing 
ratio. At weak disorder, the level statistics changes from Ginibre to Gaussian 
Orthogonal ensemble (GOE), corresponding to the spectral transition as the 
inter-chain coupling increases. The system enters into the many-body localized 
phase with increases in disorder strength and this phase possesses a real 
eigenspectrum. Our work report a spectral transition which is unique to the 
coupled fermionic chains and enhances the prospects of observing more spectral 
transitions as compared to its single-chain counterpart. We further confirm the 
MBL by eigenstate properties such as inverse participation ratio and fractal 
dimension, and show the suppression of NHSE in the localized phase. Finally, the
boundary effects and signature of MBL are corroborated in the time evolution of 
imbalance and entanglement entropy.

This work is organized as follows. Sec.~\ref{model_ham} introduces the 
interacting two-chain Hatano-Nelson model. In Sec.~\ref{results}, we discuss the
spectral transition in eigenspectrum, the level statistics, the inverse 
participation ratio, the quench dynamics of spin imbalance and entanglement 
entropy, and experimental realization of the model Hamiltonian. Finally, we 
summarize our results in Sec.~\ref{conc}.

%%%%%%%%%%%%%%%%%%%%%%%%%%%%%%%%%%%%%%%%%%%%%%%%%%%%%%%%%%%%%%%%%%%%%%%%%%%%%%
%%%%%%%%%                       Model                                    %%%%%
%%%%%%%%%%%%%%%%%%%%%%%%%%%%%%%%%%%%%%%%%%%%%%%%%%%%%%%%%%%%%%%%%%%%%%%%%%%%%%
\section{The Model}
\label{model_ham}
\begin{figure}[ht]
  \includegraphics[width=0.8\linewidth]{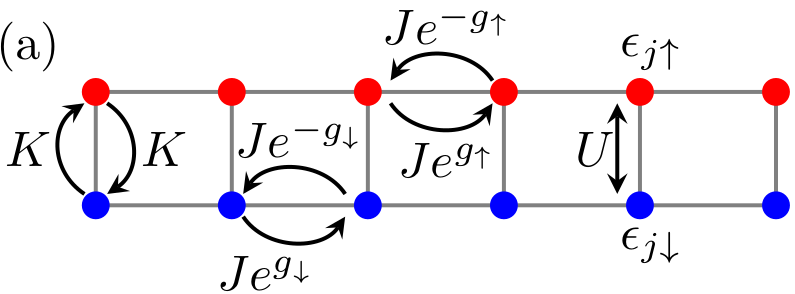}
  \includegraphics[width=\linewidth]{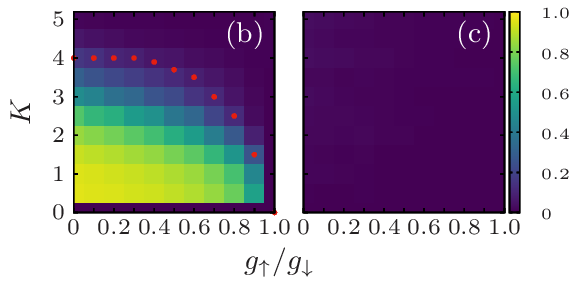}
  \caption{(a) The schematic representation of a two-chain Hatano-Nelson model.
           (b,c) The complex energy fraction $f_{\rm im}$ in 
	   $K$-$g_{\uparrow}/g_{\downarrow}$ plane. Disorder-averaged 
	   eigenspectrum distributions of a finite lattice with $L=7$ sites
           under open boundary conditions are shown for two representative cases
           : (b) $W=2$ corresponds to the ergodic regime while (c) $W=30$
           represents the MBL regime. The yellow and blue colors represent the
           parameter space of the model with complex ($f_{\rm im}=1$) and real 
	   ($f_{\rm im}=0$) eigenenergies, respectively. The red circles in (b) 
	   represent the critical inter-chain coupling strength ($K_c$) of 
	   complex-real spectral transition for different 
	   $g_{\uparrow}/g_{\downarrow}$. The interplay of $K$ and 
	   $g_{\sigma}$'s results into a (b) complex-real transitions at weak 
	   disorder and (c) at strong disorder the spectrum remains real. Here,
           we fix $U=1$ and set $J=1$ as the unit of the energy. The spectrum is
           averaged over $200$ disorder samples.}
 \label{model_spec}
\end{figure}

We consider two coupled non-Hermitian Hatano-Nelson chains of interacting 
fermions in the presence of a random potential. In the two-chain geometry, the 
two chains can be interpreted as two components of spin-$1/2$ 
fermions~[See Fig.~\ref{model_spec}(a)]. The model Hamiltonian reads as 
\begin{linenomath}
\begin{align}
  \hat{H} =&-\sum_{j,\sigma} J\left(e^{-g_{\sigma}}\hat{c}^{\dagger}_{j,\sigma} 
	     \hat{c}_{j+1,\sigma}
            + e^{g_{\sigma}}\hat{c}^{\dagger}_{j+1,\sigma} \hat{c}_{j,\sigma}
	     \right) 
	     \nonumber\\ 
           &-\sum_{j} \left(K~\hat{c}^{\dagger}_{j,\uparrow} 
	     \hat{c}_{j,\downarrow}           
	    + {\rm H.c.}\right)
             \nonumber  \\ 
	   &+ U \sum_{j} \hat{n}_{j,\uparrow} \hat{n}_{j,\downarrow} 
	    + \sum_{j,\sigma} \epsilon_{j\sigma} \hat{n}_{j,\sigma},
\label{model}
\end{align}
\end{linenomath}
where $j$ and $\sigma = \{\uparrow,\downarrow$\} represent the spatial and 
spin~(chain) indices, $J$ is the hopping strength between neighbouring lattice 
sites on the same chain, $g_{\sigma}$ is the non-Hermiticity parameter of 
$\sigma$-spin, $\hat{c}^{\dagger}_{j,\sigma} (\hat{c}_{j,\sigma})$ 
creates~(annihilates) fermion with spin $\sigma$ at $j$th site, 
$\hat{n}_{j,\sigma} = \hat{c}^{\dagger}_{j,\sigma} \hat{c}_{j,\sigma}$ is the 
occupation number operator, $K$ is the inter-chain coupling strength, $U$ is 
the onsite interaction strength, and $\epsilon_{j\sigma}$ is the random disorder
potential chosen between $[-W,W]$ with $W$ being the disorder strength. Here, 
we primarily consider the uncorrelated~(spin-dependent) disorder, which breaks 
SU(2) spin-symmetry of the model and is known to induce full localization in the 
Hermitian system~\cite{lemut_17,sroda_19,leipner_johns_19,suthar_20}. However, 
we also discuss the contrast behaviour due to spin-independent disorder 
potential~($\epsilon_{j\uparrow}=\epsilon_{j\downarrow}$) for some specific 
cases. In the present work, we set the hopping amplitude $J$ as the unit of 
energy scale, $J=1$. We consider a system of fermions at half-filling, i.e. the 
total number of fermions $N = N_{\uparrow} + N_{\downarrow} = L$ with $L$ being 
the number of lattice sites along the ladder. It is also worth noting, according 
to the symmetry class of a non-Hermitian 
system~\cite{hamazaki_20,kawabata_19,zhou_19}, the model considered here belongs 
to the symmetry class AI that preserves the time reversal symmetry~$(H=H^{*})$ 
and breaks the transposition symmetry~$(H\neq H^{T})$.

 In non-Hermitian systems with non-reciprocal hoppings, the spectra is extremely 
sensitive to the boundary conditions. For the two-chain model with dissimilar 
non-Hermiticity parameter $g_{\uparrow} \neq g_{\downarrow}$, the inter-chain 
coupling $K$ can further yield non-trivial interference between chains. When 
$K=0$, the open boundary condition allows us to remove $g_{\sigma}$ by an 
imaginary gauge transformation~(IGT)~\cite{hatano_21}, 
$c_{j,\sigma} \rightarrow e^{g_{\sigma}j} c_{j,\sigma}$ and 
$c^{\dagger}_{j,\sigma} \rightarrow e^{-g_{\sigma}j} c^{\dagger}_{j,\sigma}$. 
As a result, Eq.~\eqref{model} becomes a Hermitian disordered Hubbard model and 
all the many-body eigenenergies are real under OBC. For any $K\neq 0$, if 
$g_{\uparrow} \neq g_{\downarrow}$, $g_{\sigma}$ cannot be removed by such a 
transformation, and imaginary parts could appear in the spectrum. While the 
complex~(real) spectrum of the single Hatano-Nelson chain is unique to the 
PBC~(OBC), much richer eigenspectrum and many-body dynamics arise from the 
two-chain model under OBC. 

%%%%%%%%%%%%%%%%%%%%%%%%%%%%%%%%%%%%%%%%%%%%%%%%%%%%%%%%%%%%%%%%%%%%%%%%%%%%%%
%%%%%               Real-complex transition of eigenspectrum              %%%%
%%%%%%%%%%%%%%%%%%%%%%%%%%%%%%%%%%%%%%%%%%%%%%%%%%%%%%%%%%%%%%%%%%%%%%%%%%%%%%
\section{Results and Discussions}
\label{results}
\subsection{Real-complex transition of eigenspectrum}
\label{eigenener}
We first discuss the characteristic properties of the change in the 
eigenspectrum as a function of the inter-chain coupling $K$ and non-Hermiticity 
parameters $g_{\sigma}$. To manifest the complex spectrum, we define the 
fraction of the complex energies $f_{im}=\overline{D_{im}/D}$, where $D_{im}$ is
the number of complex eigenenergies with nonzero imaginary part, $D$ is the 
total number of eigenenergies, and the overline denotes the disorder average. 
The eigenenergies are defined as complex if $|{\rm Im}\{E\}|\geqslant C$ with a
cut-off $C=10^{-13}$, which is identified based on the machine error. Since the
model Hamiltonian preserves the time-reversal symmetry, we find that the 
imaginary parts of the energies appear symmetric to the real 
axis~[see Appendix~\ref{app_A} for details].

Under OBC, we first show the disorder-averaged fraction of the complex energies 
for weak disorder strengths $(W=2)$ in Fig.~\ref{model_spec}(b). The phase 
diagram can be divided in various regimes (i) $K=0 \rightarrow K\neq 0$, 
(ii) $g_{\uparrow}\neq g_{\downarrow} \rightarrow g_{\uparrow}=g_{\downarrow}$, 
(iii) $K \gg 1$. (i) When tuning the inter-chain coupling from decoupled ($K=0$) 
to the coupled ($K\neq 0)$ limit, we numerically verify that the phase diagram, 
except for $g_{\uparrow}/g_{\downarrow}=1$, exhibits a real-complex spectral 
transition of the many-body eigenenergies~\cite{rafi_22}. In the limit of weak 
inter-chain coupling ($K\approx10^{-5}$), the energy spectrum of the system 
resembles that of two individual uncoupled chains. It is important to note that 
a single Hatano-Nelson chain under OBC possess real spectrum, this is because 
the non-reciprocal hopping can be gauged out using IGT and a non-Hermitian 
Hamiltonian with real energies can be mapped to a Hermitian Hamiltonian. To 
elucidate the effect of nonzero $K$ on the eigenenergy distribution, we first 
consider the $g_{\uparrow} = g_{\downarrow}$ case where the forward and 
backward hoppings are identical for both of the chains. In this scenario, the 
anomalous localization due to NHSE still survives and the validity of IGT makes 
the energy spectrum real. On the other hand, the 
$g_{\uparrow}\neq g_{\downarrow}$ case leads to dissimilar inverse skin lengths 
(of NHSE) for each of the chains. The effects of $g_{\sigma}$'s can not be 
removed through IGT for unequal non-Hermiticity parameters. The collective 
effects of the two different skin modes result into a complex eigenenergy 
phase~\cite{rafi_22}. Hence, as the inter-chain coupling is varied from a 
decoupled ($K=0$) to the coupled ($K\neq 0)$ limit, we numerically find that 
the phase diagram, except for $g_{\uparrow}/g_{\downarrow}=1$, exhibits a 
real-complex spectral transition of the many-body eigenenergies. (ii) We 
further discuss the effects of $g_{\uparrow}/g_{\downarrow}$ on the 
eigenspectrum. A smooth complex-real spectral transition is observed as the 
ratio $g_{\uparrow}/g_{\downarrow}$ approaches unity. Note that the OBC 
eigenenergies are real at $g_{\uparrow}/g_{\downarrow}=1$ because IGT is valid 
for this case. As the $g_{\uparrow}/g_{\downarrow}$ ratio increases, the 
critical inter-chain coupling strength ($K_c$) of the complex-real transition 
decreases, as evident in Fig.~\ref{model_spec}(b). For a hybridized coupled 
chain with $g_{\uparrow}=0$ and $g_{\downarrow} \neq 0$ or vice versa, $K_c$ 
reaches the maximum and for $g_{\uparrow}=g_{\downarrow}$, $K_c=0$ as for the 
later case the system can be mapped to a Hermitian one. (iii) At sufficiently 
higher $K$, the effect of the non-Hermiticity is suppressed and the system 
possesses real spectra irrespective of any $g_{\uparrow}/g_{\downarrow}$. The 
quantitative variation of $f_{\rm im}$ for different system sizes is discussed 
in the Appendix~\ref{app_A}, which confirms the robustness of the phase diagram 
with $L$.

At strong disorder strengths, the system is expected to be in the many-body 
localized phase. As shown in Fig.~\ref{model_spec}(c), the OBC eigenenergies 
remain real in entire $K$-$g_{\uparrow}/g_{\downarrow}$ plane since the strong 
disorder potential destroys the non-Hermiticity. Hence, we find a 
disorder-induced complex-real transition. It is important to note that for the 
coupled-chain system this transition exists not only for OBC but also for PBC, 
in stark contrast to the occurrence of the inter-chain coupling driven 
transition which is sensitive to the boundary conditions. We have provided the 
details of the eigenspectrum with PBC in Appendix~\ref{app_A}.

%%%%%%%%%%%%%%%%%%%%%%%%%%%%%%%%%%%%%%%%%%%%%%%%%%%%%%%%%%%%%%%%%%%%%%%%%%%%%%
%%%%%                      Spectral statistics                            %%%%
%%%%%%%%%%%%%%%%%%%%%%%%%%%%%%%%%%%%%%%%%%%%%%%%%%%%%%%%%%%%%%%%%%%%%%%%%%%%%%
\subsection{Spectral statistics}
\label{spec_prop}
We now investigate the level statistics to unveil the universal features of the 
eigenenergies. For non-Hermitian disordered systems with time-reversal symmetry, 
the localized phase follows the real Poisson ensemble while the delocalized 
phase follows the Ginibre ensemble~\cite{ginibre_65,grobe_88,haake_13}. The 
nearest-level spacings for an eigenenergy $E_{i}$ are defined as 
$|E_{i} - E^{\rm NN}_{i}|$, where $E^{\rm NN}_{i}$ is an eigenvalue nearest to 
$E_{i}$ in the complex energy plane. We first perform an unfolding of the 
spectrum to obtain the histogram of the Euclidean distance between 
nearest-neighbour eigenvalues~\cite{haake_13,hamazaki_20}. Here, the unfolding 
procedure is applied to both complex and real spectra of the model. To get the 
unfolding spectrum, we first compute the nearest-neighbour distance of the 
eigenvalues
\begin{equation}
  d_{1,i} \equiv {\rm min}_{j} |E_i - E_j|.
\end{equation}
Next, the local mean density of the eigenvalues is computed as
\begin{equation}
  \bar{\rho} = \frac{n}{\pi d^{2}_{n,i}},
\end{equation}
where $n$ is sufficiently larger than unity~$(\approx 30)$, but very small 
compared to the Hamiltonian matrix size. Here, $d_{n,i}$ is the $n$th 
nearest-neighbour distance from $E_{i}$. The rescaled nearest-neighbor distance 
$s_{i}$ is obtained as
\begin{equation}
  s_{i} = d_{1,i} \sqrt{\bar{\rho}},
\end{equation}
which removes the dependence of the local density of eigenvalues on the level 
spacing. Finally, the statistics of nearest-neighbour spacings are computed 
from $s_{i}$. 
\begin{figure}[ht]
  \includegraphics[width=\linewidth]{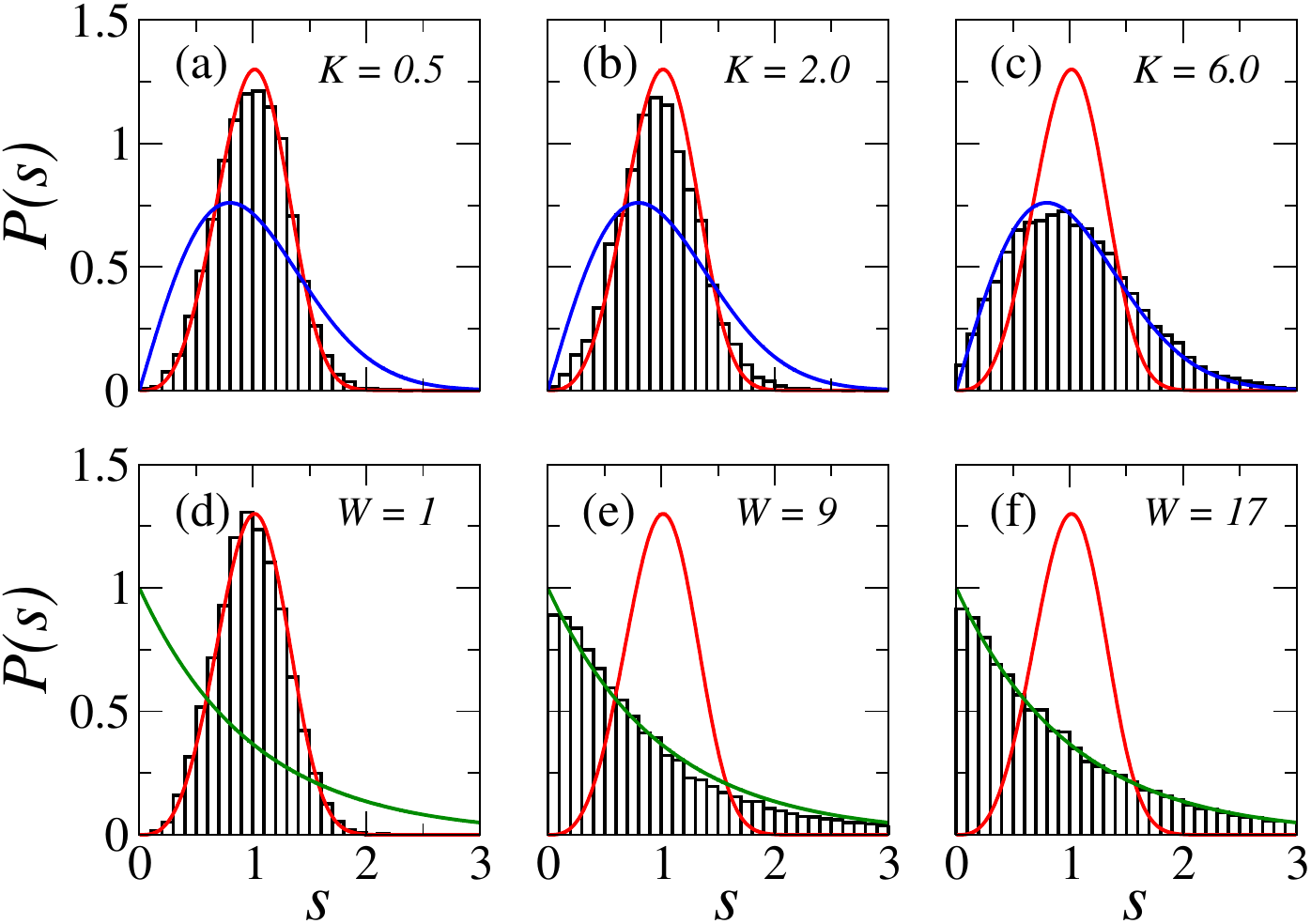}
  \caption{(a,b,c) The unfolded nearest-level spacing distributions with $K$ at 
	   $W=2$ and $g_{\uparrow}/g_{\downarrow}=0.5$. At smaller $K$ the 
	   distribution follows Ginibre distribution and at higher $K$ follows 
	   GOE distribution. (d,e,f) The level-spacing distributions as a 
	   function of the disorder strengths for $K=1$ and 
	   $g_{\uparrow}/g_{\downarrow}=0.5$. The disorder-driven complex-real 
	   transition corresponds to the Ginibre-to-Poisson level-spacing 
	   transition. The red, blue, and green lines represent the Ginibre, 
	   GOE, and~(real) Poisson distributions, respectively. Here, the 
	   system size $L=8$, onsite interaction $U=1$, and disorder 
	   average is performed for $100$ realizations.}
\label{dist}
\end{figure}

At weak disorder, the complex energy spectrum with smaller $K$ and  
$g_{\uparrow} \neq g_{\downarrow}$ obeys the Ginibre distribution 
$P^{\rm C}_{\rm Gin}(s) = cp(cs)$ which describes the ensemble of non-Hermitian
Gaussian random matrices~\cite{haake_13,hamazaki_20,prosen_20}. Here,
\begin{equation}
  p(s) = \lim_{N\rightarrow\infty}
         \left[\prod_{n=1}^{N-1}e_n(s^2)e^{-s^2}\right]
         \sum_{n=1}^{N-1}\frac{2s^{2n+1}}{n!e_n(s^2)}
\end{equation}
with $e_n(x)=\sum_{m=0}^n\frac{x^m}{m!}$ and $c=\int_0^\infty ds~s~p(s)=1.1429$
~\cite{markum_99,haake_13}. We further find that for the strong inter-chain 
coupling $K$ or equal $g_{\sigma}$'s of two chains, the real eigenspectrum of 
weak disorder case follows the level statistics of GOE. The level-spacing 
distribution of GOE is
\begin{equation}
  P^{\rm R}_{\rm GOE}(s) = \frac{\pi s}{2}~\exp(-\pi~s^2/4).
\end{equation}
The nearest-neighbour level-spacing distributions as a function of $K$ at $W=2$ 
are plotted in Fig.~\ref{dist}(a,b,c). It shows that for smaller $K$, the 
distribution is a Ginibre distribution while at larger $K$ the system follows 
GOE distribution. Hence, the general features of the non-Hermitian Hamiltonian 
with purely real eigenvalues can be mapped to a Hermitian Hamiltonian. On the 
other hand, at strong disorder case, the localized phase with real eigenspectra 
is characterized by the Poisson level distribution 
$P^{\rm R}_{\rm Po}(s) = \exp(-s)$. The level-spacing distribution as a function 
of $W$ is illustrated in Fig.~\ref{dist}(d,e,f), which suggests the MBL phase 
transition of two-chain Hatano-Nelson model.

\begin{figure}[ht]
  \includegraphics[width=\linewidth]{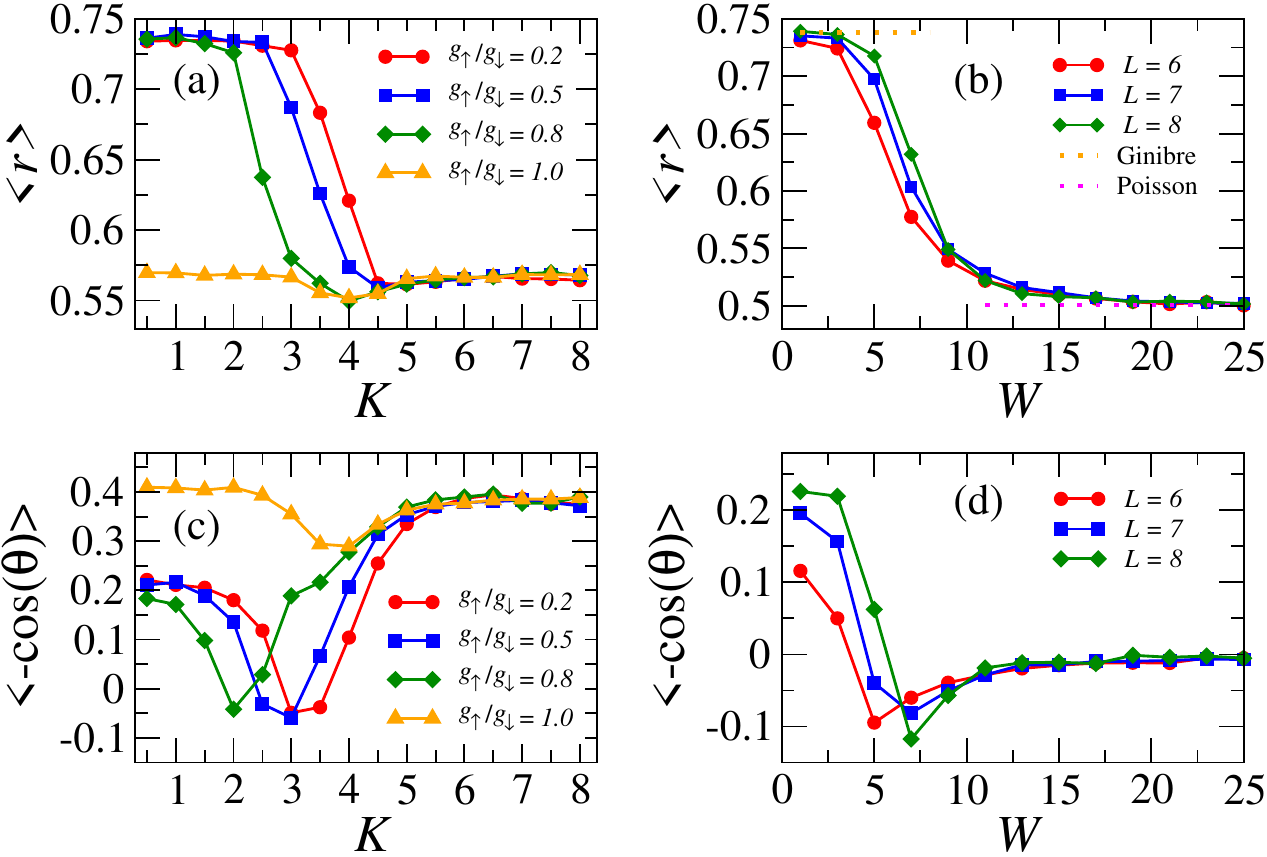}
  \caption{The average level spacing ratio $\langle r \rangle$ for the coupled
           Hatano-Nelson model exhibiting AI symmetry class. (a) Evolution of
           $\langle r \rangle$ as a function of $K$ for different
           non-Hermiticity parameters at $W=2$. As $g_{\uparrow}/g_{\downarrow}$
           ratio increases, the $K_c$ (demarcating the two level statistics)
           pushed towards lower $K$ and at $g_{\uparrow}/g_{\downarrow}=1$ the
           system exhibits Wigner-Dyson statistics. (b) Evolution of
           $\langle r \rangle$ as a function of $W$ for three system sizes at
           $K=1$ and $g_{\uparrow}/g_{\downarrow} = 0.2$. At lower $W$, the
           ergodic phase of the system (with complex eigenenergies) follows
           Ginibre ensemble statistics and at strong $W$, the MBL follows real
           Poisson statistics. The average value of $-\cos(\theta)$ as a
           function of (c) the inter-chain couplings at $W=2$ and (d) disorder
           strengths for $K=1$ and $g_{\uparrow}/g_{\downarrow}=0.2$. The system
           size is $L=8$, onsite interaction $U=1$, and disorder average is
           performed for $100$ realizations.}
\label{rbar1}
\end{figure}
We further study the complex level spacing ratio~\cite{prosen_20}. The 
level-spacing ratio is a dimensionless complex variable, 
$z_i \equiv [(E_{i} - E^{\rm NN}_{i})/(E_{i} - E^{\rm NNN}_{i})] 
\equiv r_i e^{i\theta_{i}}$ with the amplitude $ r_i \equiv|z_i|$, 
which also allows us to extract the angular information. Here, $E^{\rm NNN}_{i}$ 
is a next-nearest-neighbour eigenvalue to $E_{i}$. The mean level spacing ratio 
$\langle r \rangle$ is obtained by the average of $r_i$ over the energy window 
and number of disorder realizations. This definition is the generalization of 
the well-known gap-ratio defined for Hermitian isolated quantum 
systems~\cite{oganesyan_07,atas_13}. Here, we consider the energy window to be 
$10$\% eigenvalues around the center of the eigenspectrum in the complex energy 
plane. This allows us to obtain a large number of eigenvalues for the level 
statistics and ascertain that their eigenstates share similar localization 
properties. The number of disorder realizations is chosen such that the total 
number of eigenvalues is $\sim 10^{6}$. 

We first consider the evolution of $\langle r \rangle$ as a function of $K$ for 
weak disorder~$(W=2)$~[Fig.~\ref{rbar1}(a)]. For smaller $K$ and 
$g_{\uparrow}\neq g_{\downarrow}$, the $\langle r \rangle$ attains a constant 
value $\approx0.74$, which is for the Ginibre ensemble~\cite{prosen_20}. When 
$K$ increases to the strong coupling limit~$(K\gg1)$, Fig.~\ref{rbar1}(a) shows 
a transition to $\langle r \rangle\approx 0.56$ for the GOE 
distribution~\cite{peron_20}. This transition is consistent with the 
corresponding complex-real transition of the eigenspectrum shown in 
Fig.~\ref{model_spec}(b). In Fig.~\ref{rbar1}(a), it is evident that the $K_c$, 
which demarcates Ginibre and GOE statistics, lowers as the ratio 
$g_{\uparrow}/g_{\downarrow}$ approaches unity. For 
$g_{\uparrow}/g_{\downarrow}=1$, the $\langle r \rangle$ remains nearly GOE, 
because the non-Hermiticity can be removed by the imaginary gauge transformation
under OBC. It is interesting to note that for $g_{\uparrow}/g_{\downarrow}=1$, 
the correlated~(spin-independent) disorder potential leads to a Poisson level 
distribution due to inherent SU(2) spin symmetry. Therefore, to have GOE 
statistics, we need uncorrelated disorder potential which breaks the spin 
symmetry~[see Appendix~\ref{app_B} for details].

Fig.~\ref{rbar1}(b) shows $\langle r \rangle$ as a function of the disorder 
strength $W$  for weak inter-chain coupling~($K=1$).  We demonstrate that the 
disorder-induced complex-real transition is accompanied by a change in 
$\langle r \rangle$ from Ginibre to real Poisson statistics. It is important to 
stress that the present non-Hermitian model has $\langle r \rangle \approx 0.5$ 
for the real Poisson statistics, which is different from the conventional 
$\langle r \rangle \approx 0.38$ of Hermitian many-body 
systems~\cite{oganesyan_07,atas_13}, even both being characterized for real 
spectra. This fact is consistent with a recent study which characterizes the 
spacing ratio for the spectra of non-Hermitian random matrices~\cite{peron_20}. 
Furthermore, the critical disorder strength of the transition $W_c \approx 9$ 
can be obtained by inspecting the crossing of $\langle r \rangle$-curves for two 
largest systems. Combining Fig.~\ref{rbar1} (a) and (b), we note that for strong 
inter-chain coupling or $g_{\uparrow}/g_{\downarrow}=1$, the disorder leads to a 
transition from GOE to Poisson distribution. Thus, at strong $W$, the effects of 
the disorder potential prevails and $\langle r \rangle$-value reaches a 
stationary value of the real Poisson statistics.   

 Likewise, the single-number signature of $\langle -\cos(\theta) \rangle$ 
distinguishes the ergodic and localized phase~\cite{prosen_20}. In 
Fig.~\ref{rbar1}(c) and Fig.~\ref{rbar1}(d), we show the evolution of the 
disorder-averaged mean $-\cos(\theta)$ with $K$ and $W$. We first consider 
$\langle -\cos(\theta) \rangle$ as a function of $K$ for weak 
disorder~[Fig.~\ref{rbar1}(c)]. At small $K$, we numerically find 
$\langle-\cos(\theta) \rangle\approx 0.22$, that corresponds to the Ginibre 
level distribution. This value is consistent to the analysis for single 
Hatano-Nelson chain~\cite{prosen_20}. As $K$ increases, we find a dip in 
$\langle-\cos(\theta) \rangle$ which demarcates the complex- and real-energy 
phases. Moreover,  for larger $g_{\uparrow}/g_{\downarrow}$ 
(with $g_{\uparrow} \neq g_{\downarrow}$) the dip occurs at smaller $K$. On the 
other hand, for $g_{\uparrow} /g_{\downarrow}=1$ the distribution of 
$\langle-\cos(\theta) \rangle$ does not change sign. The main features of 
$\langle-\cos(\theta) \rangle$ are in consonance with the eigenspectrum 
transition and evolution of $\langle r \rangle.$ In Fig.~\ref{rbar1}(d), we 
further show the variation of $\langle -\cos(\theta) \rangle$ as a function of 
$W$ for $K=1$ and $g_{\uparrow}/g_{\downarrow}=0.2$. As $W$ increases, the 
value of $\langle -\cos(\theta) \rangle$ decreases, becomes negative, and 
eventually reaches $\langle -\cos(\theta) \rangle = 0$ for the Poisson 
statistics at strong disorder. This again confirms the disorder-induced 
spectral and localization transition.

%%%%%%%%%%%%%%%%%%%%%%%%%%%%%%%%%%%%%%%%%%%%%%%%%%%%%%%%%%%%%%%%%%%%%%%%%%%%%%
%%%%%                      Inverse Participation Ratio                    %%%%
%%%%%%%%%%%%%%%%%%%%%%%%%%%%%%%%%%%%%%%%%%%%%%%%%%%%%%%%%%%%%%%%%%%%%%%%%%%%%%
\subsection{Inverse Participation Ratio}
\label{ipr_prop}
 To further understand the interplay between localization and NHSE in real 
space, we characterize the wavefunctions by employing the inverse participation 
ratio~(IPR) and fractal dimension~(FD). For non-Hermitian systems, the IPR can 
be defined in two ways: one using $n$th left or right eigenstates and other is 
defined under biorthogonal basis from both left and right eigenstates. These 
are defined as 
\begin{eqnarray}
  I_{n,s} = \frac{\sum_{i,\sigma} |\psi^{n}_{i,\sigma}|^{2}}{(\sum_{i,\sigma} 
                                  |\psi^{n}_{i,\sigma}|)^{2}}; ~~~
  I_{nB,s} = \frac{\sum_{i,\sigma} |\tilde{\psi}^{n}_{i,\sigma}|^{2}}
	                           {(\sum_{i,\sigma} 
                                   |\tilde{\psi}^{n}_{i,\sigma}|)^{2}},
\end{eqnarray}
where $s$ is the disorder realization, the subscript $nB$ represents the 
biorthogonal IPR, $\psi^{n}_{i,\sigma} \equiv (\langle n| b_{i,\sigma} 
\rangle^{*}) \langle n| b_{i,\sigma} \rangle$ and 
$\tilde{\psi}^{n}_{i,\sigma}\equiv(\langle \tilde{n}| b_{i,\sigma} \rangle^{*}) 
\langle n| b_{i,\sigma} \rangle$ with $n$ and $\tilde{n}$ are the right and 
corresponding left eigenstates, and $|b_{i,\sigma}\rangle$ are Fock space 
chosen as computational basis. The IPRs are further first averaged over 
numerous disorder realizations and then over the spectrum to get the mean IPRs, 
$I_{\rm avg} = \sum_{n} \overline{I}_{n}/D$ and 
$I^{B}_{\rm avg} = \sum_{n} \overline{I}_{nB}/D$ with $D$ being the dimension 
of the Hilbert space and overline denotes the disorder average. The 
disorder-averaged IPRs of eigenstate $n$ are shown in Appendix~\ref{app_C}. For 
delocalized states, the IPR approaches zero in the thermodynamic limit while 
for localized states it saturates to a finite value $(\approx 1)$. The fractal 
dimension of an eigenstate is another measure which is recently devised to 
examine the localization properties of many-body 
systems~\cite{mace_19,backer_19}. The mean FD can be directly constructed from 
the mean IPRs as $\eta = -\ln(I_{\rm avg})/\ln D$ and biorthogonal FD 
$\eta_{B} = -\ln(I^B_{\rm avg})/\ln D$. The extended and localized phases are 
recognized by $\eta \rightarrow 1$ and $\eta \rightarrow 0$, respectively. 
\begin{figure}[ht]
  \includegraphics[width=\linewidth]{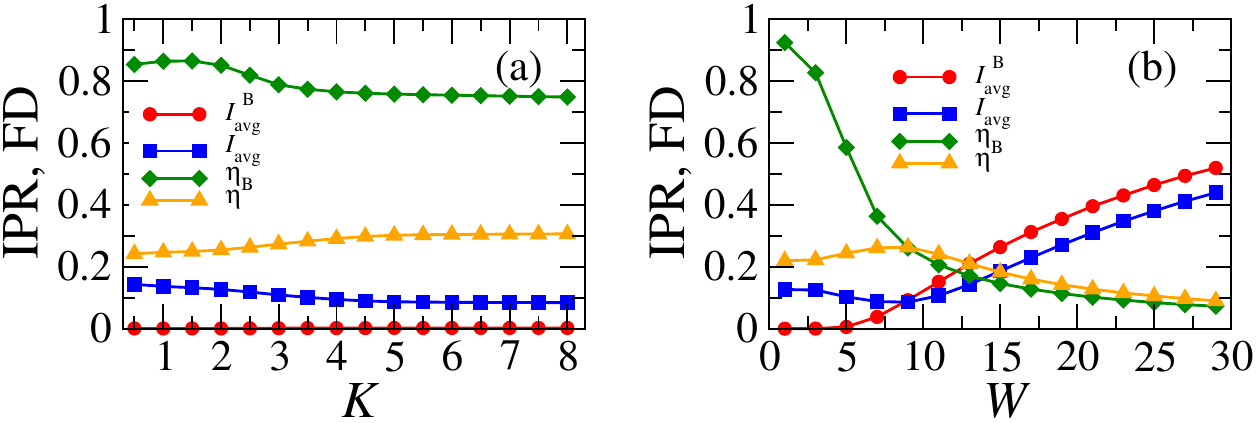}
  \caption{Disorder-averaged IPR and FD of the right eigenstates 
	   (squares,triangles) and biorthogonal eigenstates (circles,diamonds) 
	   for the system size $L=8$ and $g_{\uparrow}/g_{\downarrow}=0.5$ at 
	   (a) $W=2$ and (b) $K=1$. Here, the disorder average is performed for 
	   $100$ samples.}
\label{ipr_fd}
\end{figure}

In Fig.~\ref{ipr_fd}(a) we show the mean IPR and FD~(defined in both ways) of 
open chains as a function of $K$ for $W=2$.  It is interesting to note that the 
value of $I_{\rm avg}$ is nonzero even at weak disorder, whereas 
$I^{B}_{\rm avg}$ is zero. This suggests that the finite $I_{\rm avg}$ is due 
to the NHSE which plays a significant role at weak disorder, when the 
biorthogonal density distributions do not suffer from the NHSE. This fact can 
be understood from the similarity transformation $\ket{n'} = S^{-1} \ket{n}$ 
and $\bra{n'} = \bra{\tilde{n}}S$, with $\ket{n'}$ being the eigenstate of the 
corresponding Hermitian system. Hence, the biorthogonal density distributions 
are devoid of the NHSE. It is important to note that the choice of eigenstates 
for IPR is only important for open chains as under PBC the NHSE is absent. 
Furthermore, the corresponding contrast behaviour in $\eta$ and $\eta_{B}$ is 
also noted, cf.~Fig.~\ref{ipr_fd}(a). The evolution of $I^{B}_{\rm avg}$ and 
$\eta_{B}$ with $K$ confirms the delocalization at $W=2$. While the system 
possesses a complex-real spectral transition with increase in $K$, as shown in 
Fig.~\ref{model_spec}(b), this transition does not overlap with the 
localization transition. 

We further present the evolution of mean IPR and FD with $W$ in 
Fig.~\ref{ipr_fd}(b). At lower $W$, the $I^{B}_{\rm avg}$ is small but 
$I_{\rm avg}$ acquires a finite value due to the interference of the NHSE. The 
$\eta_{B}$ approaches unity at lower $W$ representing the delocalization. The 
value of mean IPR~(FD) increases~(decreases) with $W$, confirms the 
disorder-driven localization in the non-Hermitian two-chain model. The observed 
behaviours of IPR and FD show that the disorder-driven localization transition 
overlaps to the complex-real spectral transition. In addition, we note that in 
contrast to the weak-disorder regime where a significant difference between 
$\eta$ and $\eta_{B}$ exists, at strong disorder the values predicted by 
$\eta$ and $\eta_{B}$ coincide. This indicates the suppression of NHSE in the 
presence of strong disorder strength. The finite-size effects of IPR is 
discussed in Appendix~\ref{app_C}.

%%%%%%%%%%%%%%%%%%%%%%%%%%%%%%%%%%%%%%%%%%%%%%%%%%%%%%%%%%%%%%%%%%%%%%%%%%%%%%
%%%%                       Dynamical Properties                           %%%%
%%%%%%%%%%%%%%%%%%%%%%%%%%%%%%%%%%%%%%%%%%%%%%%%%%%%%%%%%%%%%%%%%%%%%%%%%%%%%%
\subsection{Dynamical Properties}
\label{dyna_prop}
Here we study the non-equilibrium time dynamics of the non-Hermitian system 
from the perspective of quantum trajectories with no-jump condition for the 
continuously measured system~\cite{daley_14,hamazaki_19}. By choosing an 
arbitrary initial state $\left|\psi_{0}\right\rangle$ at $t=0$, the time 
dynamics is encoded in the wavefunction
\begin{equation}
  \ket{\psi_{t}} = \frac{e^{-{i}\mathcal{H}t/\hbar}\ket{\psi_{0}}}
                   {\sqrt{\bra{\psi_{0}}
		   e^{{i}\mathcal{H}^{\dagger}t/\hbar}
		   e^{-{i}\mathcal{H}t/\hbar}\ket{\psi_{0}}}},
\label{wavefunction}
\end{equation}
which is governed solely by the non-Hermitian effective 
Hamiltonian~$\mathcal{H}$. With this time-dependent wavefunction, all the 
dynamical properties can be explored. 
\begin{figure}[ht]
  \includegraphics[width=\linewidth]{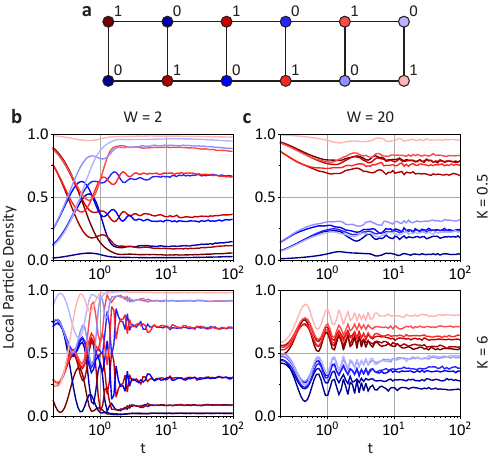}
  \caption{(a) Schematic of the initial state $\hat{c}_{1,\uparrow}^{\dagger}
           \hat{c}_{2,\downarrow}^{\dagger}\hat{c}_{3,\uparrow}^{\dagger}
           \hat{c}_{4,\downarrow}^{\dagger}\cdots\bigl|vac\bigr\rangle$.
           The red (blue) dots are initially occupied  (unoccupied) sites, and
           the dark to light gradients denote the spatial position from left to
           right. (b,c) The time evolution of the local particle density at each 
	   site for weak~($K=0.5$, upper panel) and strong~($K=6$, lower panel)
	   inter-chain couplings. The dynamics at weak~($W=2$) and 
	   strong~($W=20$) disorder strengths are shown in (b) and (c), 
	   respectively. The plots are obtained by averaging over $500$ disorder
	   realizations, and the other parameters are $U=1$ and 
	   $g_{\uparrow}/g_{\downarrow}=0.5$.}
\label{localparticledensity}
\end{figure}

We first discuss the dynamics of the local particle density $n_{j,\sigma}(t)$ 
and spin imbalance $I_{s}(t)$ for different inter-chain couplings $K$ and 
disorder strengths $W$. Here the spin imbalance is defined as
\begin{equation}
  {I}_{s} (t) = \frac{1}{L}\sum_{j=1}^{L}(-1)^{j-1}
	                   (n_{j,\uparrow}(t)-n_{j,\downarrow}(t)),
\label{imbalance}
\end{equation}
whose long-time stationary value effectively serves as an order parameter of 
many-body localized phase. We choose the N\'eel ordered state 
$\ket{\uparrow\downarrow\uparrow\downarrow\cdots}$ as an initial state, whose 
schematic representation is shown in Fig.~\ref{localparticledensity}(a). This 
state has ${ I}_{s}(0)=1$ at initial time~($t=0$). 

\begin{figure}[ht]
  \includegraphics[width=\linewidth]{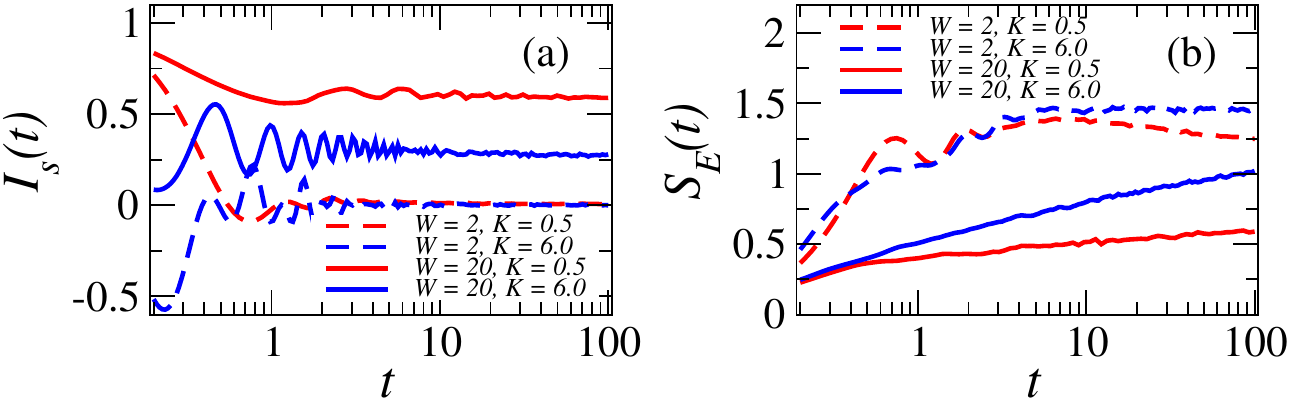}
  \caption{The time dynamics of (a) spin imbalance and (b) bipartite 
	   entanglement entropy for $W=2$ (dashed-line) and $W=20$ (solid-line) 
	   at $K=0.5$ (blue) and $K=6$ (red). At $W=20$, a non-vanishing 
	   stationary value of ${I}_{s}(t)$ and the logarithmic growth of 
	   $S_{E}(t)$ characterize the MBL. At weak disorder, the long-time 
	   dynamics of the entanglement entropy exhibits the decrease in the 
	   complex-energy phase~($K=0.5$) while remains steady in the 
	   real-energy phase~($K=6$). The disorder average is performed over 
	   $500$ realizations. Here, the onsite interaction strength $U=1$ and 
	   $g_{\uparrow}/g_{\downarrow}=0.5$.}
\label{imbal_ent}
\end{figure}

At weak disorder~($W=2$), the time evolution of the local particle density for 
both small and strong inter-chain couplings are shown in 
Fig.~\ref{localparticledensity}(b). We observe that the saturation values of 
the local particle densities for $\ket{\uparrow}$ and $\ket{\downarrow}$ 
coincide at the same spatial indices. Thus,  the corresponding ${I}_{s}(t)$ in 
Fig.~\ref{imbal_ent}(a) relaxes to zero as time evolves, losing memory of 
initial ordering, suggests delocalization of the system. On the other hand, 
Fig.~\ref{localparticledensity}(c) for strong disorder~($W=20$) shows that the 
stationary values of the particles densities for $\ket{\uparrow}$ and 
$\ket{\downarrow}$ do not coincide at the same spatial indices, which leads to 
a non-vanishing steady value of ${I}_{s}(t)$ at long times in 
Fig.~\ref{imbal_ent}(a). This initial-state memory retention indicates the 
disorder-driven MBL. In addition, we find that at strong disorder, larger $K$ 
suppresses the steady value of ${ I}_{s}(t)$. This is because the $K$ term 
couples different spin sectors and scrambles the initial spin ordering. 
Therefore, as the inter-chain coupling $K$ increases, the disorder-driven 
localization transition occurs at higher $W$. 

It is interesting to note that although the system is extended in character at 
weak disorder, the larger local particle density appears at the site closer to 
the right end of the chain [Fig.~\ref{localparticledensity}(b)]. This anomalous 
localization stems from the non-reciprocal tunnelings. However, we stress that 
this localization is suppressed by strong disorder in the MBL 
phase~[Fig.~\ref{localparticledensity}(c)]. It is worth mentioning that the 
suppression of the NHSE-induced localization at strong disorder is also observed 
in the eigenstate properties, as discussed in subsection~\ref{ipr_prop}. 

We further investigate the dynamics of half-chain von Neumann entanglement 
entropy which is defined as 
\begin{equation}
  S_{E}(t) = -\overline{\Tr [\rho_{A}(t)\ln\rho_{A}(t)]}, 
\end{equation}
where the two subsystems are denoted as $A$ and $B$ with 
$\rho_{A}(t)=\Tr_{B}[\ket{\psi_{t}}\bra{\psi_{t}}]$ being the reduced density 
matrix of subsystem $A$. Here $\Tr_B$ is the trace over degrees of freedom of 
subsystem $B$. The time evolution of $S_{E}(t)$ for various limits of $K$ and 
$W$ is shown in Fig.~\ref{imbal_ent}(b). For strong disorder strength, we find 
the long-time evolution of $S_{E}(t)$ exhibits a logarithmic growth, which is 
reminiscent of non-ergodicity in the Hermitian many-body 
localization~\cite{znidaric_08,bardarson_12}. For weak disorder, however, 
$S_{E}(t)$ can decrease after $t\approx5$ in the complex eigenenergy 
phase~($K=0.5$), but remains steady in the real eigenenergy one~($K=6$). This 
long-time behavior of $S_{E}(t)$ signifies the real-complex transition unique 
to open non-Hermitian chains. 

%%%%%%%%%%%%%%%%%%%%%%%%%%%%%%%%%%%%%%%%%%%%%%%%%%%%%%%%%%%%%%%%%%%%%%%%%%%%%%
%%%%%                     Experimental realization                         %%%%%
%%%%%%%%%%%%%%%%%%%%%%%%%%%%%%%%%%%%%%%%%%%%%%%%%%%%%%%%%%%%%%%%%%%%%%%%%%%%%%
\subsection{Experimental realization}
  The spectral and localization transitions reported in the present work can be 
qualitatively realized in two-chain fermionic lattice with asymmetric hoppings. 
Recently, it has been proposed that such non-reciprocal hopping can be 
effectively implemented by a reservoir 
engineering~\cite{poyatos_96,diehl_08,diehl_11}. It is worth mentioning that 
the implementation of a single Hatano-Nelson chain in cold-atom experiments has 
already been proposed~\cite{gong_18}. Using similar strategy in the 
implementation of a two-component fermionic systems (such as a non-Hermitian 
Su-Schrieffer-Heeger model or two-chain system considered here) and allowing a 
Rabi-coupling $K$ is feasible for cold-atoms trapped in optical 
lattices~\cite{li_19,song_19,lapp_19,li_20c,longhi_22,ren_22}. The model 
Hamiltonian of the present work can be mapped to an effective non-Hermitian 
dissipative model. In particular, the Hamiltonian Eq.~(\ref{model}) can be 
re-written into Hermitian $\hat{H}_{1} = (\hat{H} + \hat{H}^{\dagger})/2$ and 
anti-Hermitian $\hat{H}_{2} = (\hat{H} - \hat{H}^{\dagger})/2$ parts 
\begin{linenomath}
\begin{subequations}
\begin{align}
  \hat{H}_1 =&-\frac{J(e^{g_{\sigma}} + e^{-g_{\sigma}})}{2} \sum_{j,\sigma} 
	       \left(\hat{c}^{\dagger}_{j+1,\sigma} \hat{c}_{j,\sigma} 
	      +\hat{c}^{\dagger}_{j,\sigma} \hat{c}_{j+1,\sigma}\right)
               \nonumber\\
             &-\sum_{j} K \left(\hat{c}^{\dagger}_{j,\uparrow} 
	       \hat{c}_{j,\downarrow}
              +\hat{c}^{\dagger}_{j,\downarrow} \hat{c}_{j,\uparrow}\right)
               \nonumber  \\
             &+ U \sum_{j} \hat{n}_{j,\uparrow} \hat{n}_{j,\downarrow}
              + \sum_{j,\sigma} \epsilon_{j\sigma} \hat{n}_{j,\sigma}, \\
  \hat{H}_2 =&-\frac{J(e^{g_{\sigma}} - e^{-g_{\sigma}})}{2} \sum_{j,\sigma}
	       \left(\hat{c}^{\dagger}_{j+1,\sigma} \hat{c}_{j,\sigma} 
	      -\hat{c}^{\dagger}_{j,\sigma} \hat{c}_{j+1,\sigma}\right).
\end{align}
\label{h_nh_part}
\end{subequations}
\end{linenomath}
The Hermitian part of the Hamiltonian $\hat{H}_1$ in Eq.~(\ref{h_nh_part}a) with
random disorder potential can be constructed by superimposing an optical speckle
field onto an optical lattice~\cite{pasienski_10,choi_16}. The disorder strength
is proportional to the speckle field strength and therefore can be tuned. 
Moreover, with current advancement of ultracold atom experiments, it is possible 
to generate spin-dependent disorder potential by using laser beams of different 
polarizations. The onsite interaction strength is tunable by adjusting the 
$s$-wave scattering length between two fermionic components using a Feshbach 
resonance~\cite{guan_13}. Note that the implementation of a two-leg ladder 
systems in a clean system is already performed in recent 
experiments~\cite{kang_18,han_19,kang_20,ho_22,lauria_22}, where the inter-chain 
tunneling of atoms can be controlled by the Raman transitions between the 
states.

The anti-Hermitian part $\hat{H}_2$ can be implemented by considering the jump 
operator that includes a collective one-body loss~\cite{gong_18}
\begin{equation}
  \hat{L}_{j,\sigma} = \sqrt{J~|e^{g_\sigma}-e^{-g_\sigma}|}
		       \big[\hat{c}_{j,\sigma}+~i~\text{sgn}(g_\sigma) 
		       \hat{c}_{j+1,\sigma}\big],
\end{equation}
where $\text{sgn}(g_\sigma)$ indicates the sign of $g_\sigma$ and controls the 
direction of non-reciprocal tunnelings. Under no-jump condition or 
post-selection~\cite{daley_14,hamazaki_19}, the dynamics of the density matrix 
is solely governed by the following effective non-Hermitian Hamiltonian
\begin{linenomath}
\begin{align}
  \hat{H}_\text{eff} &= \hat{H}_1-\frac{i}{2}
	                \sum_{\sigma}\sum_{j=0}^{L}\hat{L}_{j,\sigma}^{\dagger}
			\hat{L}_{j,\sigma}\nonumber \\
	             &= \hat{H}_1+\hat{H}_2-~i~
		        \sum_{\sigma}\sum_{j=1}^{L} J~|\sinh g_\sigma| 
			\hat{c}_{j,\sigma}^{\dagger}\hat{c}_{j,\sigma},
\end{align}
\end{linenomath}
where the last term represents the onsite atom decay and we have considered 
the open boundary conditions at $j=0$ with 
$\hat{L}_{0,\sigma}=\sqrt{J~|e^{g_\sigma}-e^{-g_\sigma}|}
{i}~\text{sgn}(g_\sigma) \hat{c}_{1,\sigma}$ and $j=L$ with 
$\hat{L}_{L,\sigma}=\sqrt{J~|e^{g_\sigma}-e^{-g_\sigma}|}\hat{c}_{L,\sigma}$. 
It is evident that two chains have different decay terms at 
$g_\uparrow\neq g_\downarrow$. To compensate for this discrepancy, two chains 
should couple to different reservoirs to have the same onsite decay rate. In 
this case, the final effective non-Hermitian Hamiltonian differs from our model 
Hamiltonian in Eq.~(\ref{model}) by an overall decay term. Nevertheless, it 
would not affect the dynamics of the system as long as the no-jump condition 
holds, or the post-selection is considered, where the wavefunction is given by 
Eq.~(\ref{wavefunction}). In experiments such novel nonlocal loss can be 
engineered by nonlocal Rabi coupling as recently proposed in 
Ref.~\cite{gong_18}. A non-reciprocal hopping effectively creates imaginary 
gauge fields and induces a non-Hermitian Aharonov-Bohm effect. More recently, 
the unique signatures of NHSE in many-body systems are observed~\cite{liang_22}, 
the implementation of a random disorder potential in such experimental settings 
could be a possible step towards realization of the model of present study. We 
believe our results are within reach of current experimental progress and 
techniques. 

%%%%%%%%%%%%%%%%%%%%%%%%%%%%%%%%%%%%%%%%%%%%%%%%%%%%%%%%%%%%%%%%%%%%%%%%%%%%%%
%%%%                          Conclusions                                %%%%%
%%%%%%%%%%%%%%%%%%%%%%%%%%%%%%%%%%%%%%%%%%%%%%%%%%%%%%%%%%%%%%%%%%%%%%%%%%%%%%
\section{Conclusions}
\label{conc}
We discussed the eigenspectrum, level statistics, and localization properties 
of a two coupled non-Hermitian chains. We unveiled the occurrence of two 
complex-real spectral transitions. One transition is induced by the interplay 
of non-reciprocal hoppings and inter-chain coupling which is absent under 
periodic boundary conditions, while the other one is driven by a random disorder
potential and present under both open and periodic boundaries. Furthermore, we 
have studied both spectral transitions using level statistics, inverse 
participation ratio, and fractal dimension. We have shown the suppression of 
the non-Hermitian skin effect in the many-body localized phase. Finally, the 
time evolution of the local particle density, spin imbalance, and entanglement 
entropy corroborates the pivotal role of boundaries and characterizes the 
many-body localization. We believe the characteristics and results of our 
coupled-chain model are timely and pertinent, and can be readily implemented in 
various non-Hermitian systems. Our findings pave a way to further investigate 
the interplay of the boundary conditions, the disorder-driven localization, and 
the many-body dynamics in other non-Hermitian systems.

%%%%%%%%%%%%%%%%%%%%%%%%%%%%%%%%%%%%%%%%%%%%%%%%%%%%%%%%%%%%%%%%%%%%%%%%%%%%%%
%%%%                        Acknowledgments                              %%%%%
%%%%%%%%%%%%%%%%%%%%%%%%%%%%%%%%%%%%%%%%%%%%%%%%%%%%%%%%%%%%%%%%%%%%%%%%%%%%%%
\begin{acknowledgments}
  We acknowledge the support of High Performance Computing Cluster at IAMS, 
Academia Sinica. K.S. acknowledges the support by Academia Sinica. This 
research has been supported by the Ministry of Science and Technology (MOST),
Taiwan, under the Grant No. MOST-109-2112-M-001-035-MY3 (Y.-C.W. and H.H.J.),
No. MOST-110-2112-M-003-008-MY3 (Y.-C.W. and J.-S.Y.), and 
No. MOST-111-2636-M-007-009 (Y.-P.H.). Y.-P.H., H.H.J., and J.-S.Y. are 
also grateful for support from National Center for Theoretical Sciences in 
Taiwan.
\end{acknowledgments}

%%%%%%%%%%%%%%%%%%%%%%%%%%%%%%%%%%%%%%%%%%%%%%%%%%%%%%%%%%%%%%%%%%%%%%%%%%%%%%%
%%%%                         Appendix                                      %%%%
%%%%%%%%%%%%%%%%%%%%%%%%%%%%%%%%%%%%%%%%%%%%%%%%%%%%%%%%%%%%%%%%%%%%%%%%%%%%%%%
%%%% Appendix A : Complex-real transition of eigenspectrum and             %%%%
%%%%                           boundary conditions                         %%%%
%%%%%%%%%%%%%%%%%%%%%%%%%%%%%%%%%%%%%%%%%%%%%%%%%%%%%%%%%%%%%%%%%%%%%%%%%%%%%%%
\appendix
\section{Finite-size effects of complex-energy fraction and role of boundary 
         conditions}
\label{app_A}
\begin{figure}[ht]
  \includegraphics[width=\linewidth]{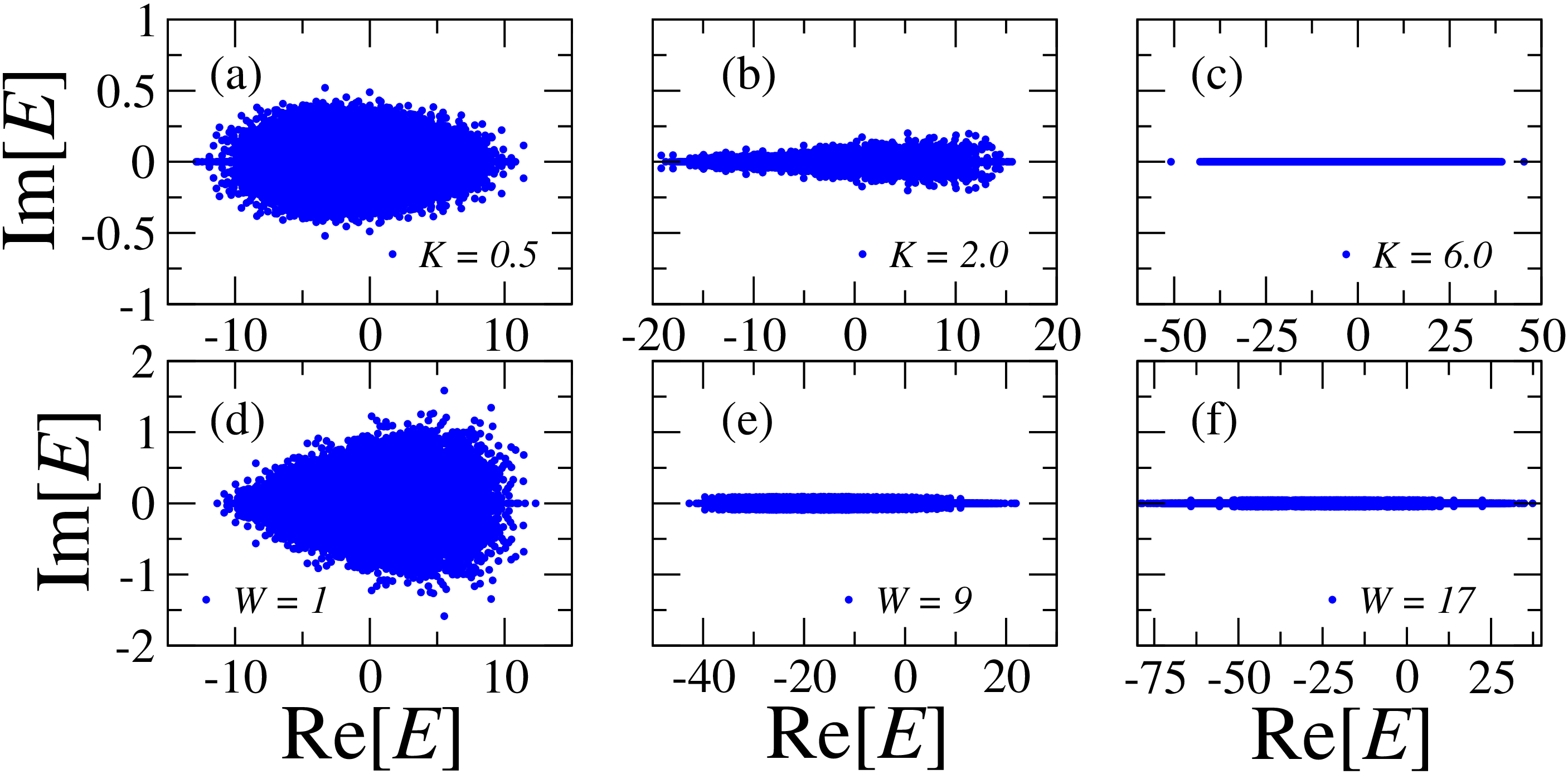} \\ \vspace{0.3cm}
  \includegraphics[width=\linewidth]{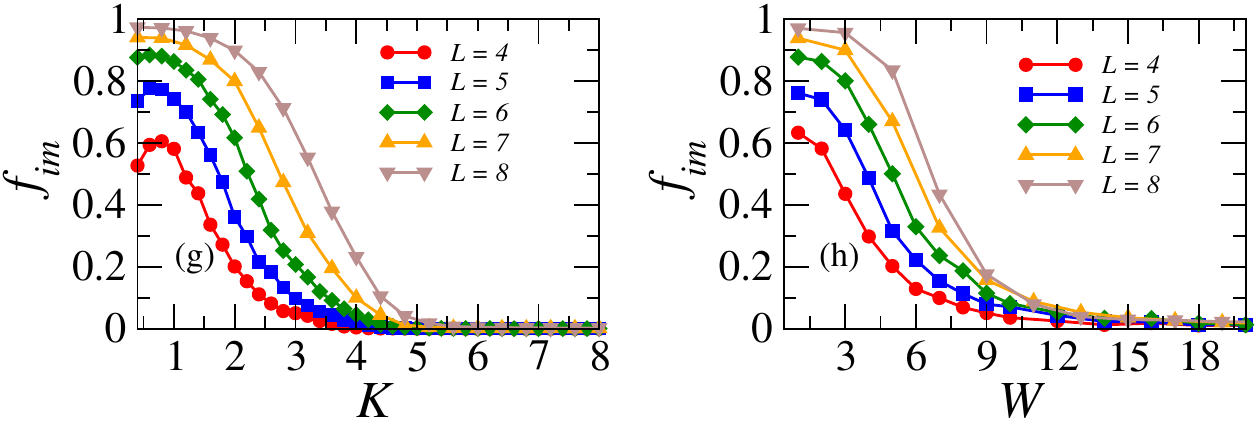}
  \caption{Eigenenergies of the non-Hermitian coupled Hatano-Nelson open chain
           for $L=8$ as a function of (a,b,c) the inter-chain coupling $K$ at 
	   $W=2$ and (d,e,f) the disorder strength $W$ at $K=1$. These are 
	   shown corresponding to Fig.~\ref{dist}(a-f) of the main text. The 
	   decrease in the fraction of complex energies with $K$ and $W$ are 
	   evident. (g,h) The disorder-averaged complex energy fraction 
	   $f_{\rm im}$ as a function of (g) $K$ and (h) $W$ for different 
	   system sizes and $g_{\uparrow}/g_{\downarrow} = 0.2$.}
\label{ener_K_W_L}
\end{figure}
The eigenenergies of the model Hamiltonian~[Eq.~\eqref{model}] for $L=8$ under 
OBC are shown in Fig.~\ref{ener_K_W_L}(a-f). Since the system respects 
time-reversal symmetry and belongs to the symmetry class AI, the complex 
spectrum is symmetric with respect to the real axis. With the increase in 
$K$~[Fig.~\ref{ener_K_W_L}(a,b,c)] and $W$~[Fig.~\ref{ener_K_W_L}(d,e,f)], the 
imaginary parts of the energies are suppressed. The complex-real transition 
exists for $g_{\uparrow}\neq g_{\downarrow}$ in a coupled non-Hermitian chain. 
For $g_{\uparrow} = g_{\downarrow}$, however, the spectrum remains real, because 
the imaginary gauge transformation is valid for this special case.

We present the finite-size effects on the complex energy fraction $f_{\rm im}$. 
Our analysis is restricted to the system size amenable to the exact 
diagonalization. We estimate the finite-size effects by comparison of 
$f_{\rm im}$ for smaller system sizes. We first discuss the inter-chain coupling 
driven spectral transition under OBC at weak disorder. The evolution of 
$f_{\rm im}$ with $K$ at $g_{\uparrow}/g_{\downarrow} = 0.2$ and $W=2$ for 
different system sizes are shown in Fig.~\ref{ener_K_W_L}(g). While the 
eigenenergies of a single Hatano-Nelson open chain are real, an infinitesimal 
coupling $K$ between the chains with $g_{\uparrow}\neq g_{\downarrow}$ leads to 
complex energies. In this analysis, we have varied the value of $K$ from a 
small ($K=0.3$) to a strong ($K=8$) coupling limit. As seen from 
Fig.~\ref{ener_K_W_L}(g), the $f_{\rm im}$ shifts to a larger value as $L$ 
increases, and at $L=8$ the fraction of the complex energies 
$f_{\rm im}\approx 1$. This suggests the stability of the complex energy phase 
in the thermodynamic limit for small $K$. On the other hand, at strong 
inter-chain coupling, the $f_{\rm im}$ converges to zero as $L$ increases. This 
shows the robustness of the spectral transitions shown in 
Fig.~\ref{model_spec}(b) of the main text. We believe the qualitative features 
of the spectral transitions holds in the thermodynamic limit, however the 
critical inter-chain coupling strength $K_c$ might vary for larger system sizes.
\begin{figure}[ht]
  \includegraphics[width=\linewidth]{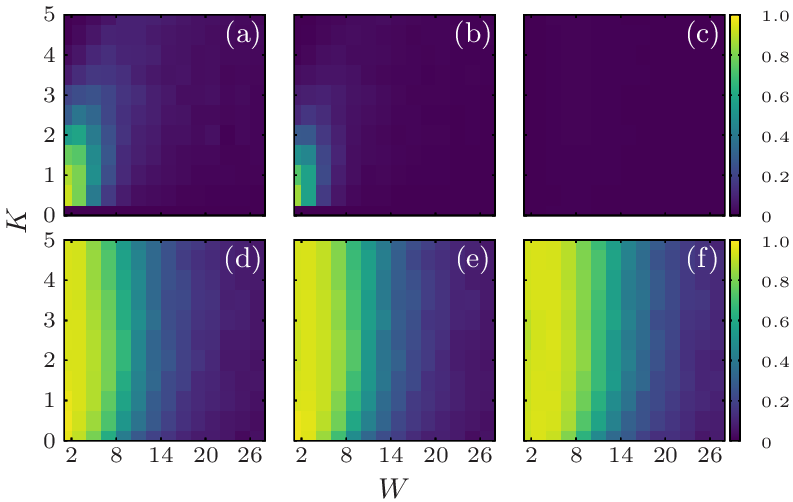}
  \caption{The disorder-averaged $f_{\rm im}$ of open coupled chains are shown 
	   for (a) $g_{\uparrow}/g_{\downarrow} = 0.2$,
           (b) $g_{\uparrow}/g_{\downarrow} = 0.6$,
           (c) $g_{\uparrow}/g_{\downarrow} = 1.0$.
           For the same values, the disorder-averaged $f_{\rm im}$ under PBC 
	   are shown in the lower panel (d,e,f). Under OBC, the
           $g_{\uparrow}/g_{\downarrow} = 1.0$ case (c) can be gauged out using
           IGT and hence the spectrum remains real in entire $K$-$W$ plane. For
           open non-Hermitian coupled-chain system, the inter-chain coupling
           also drives the system into real eigenspectrum. Here, the
           eigenspectrum is averaged over $500$ disorder realizations.}
\label{eigen_w}
\end{figure}

We further discuss the finite-size effects on the spectral transition due to 
the disorder potential. The $f_{\rm im}$ for different system sizes as a 
function of $W$ is shown in Fig.~\ref{ener_K_W_L}(h). Here, we consider $K=1$ 
and $g_{\uparrow}/g_{\downarrow} = 0.2$. For the considered parameters, at weak 
disorder strengths, the system possesses complex energies and remains in the 
delocalized phase. As disorder strength increases, we find a complex-real 
spectral transition beyond a critical value of $W$. It is important to note 
that a similar transition also appears in a single Hatano-Nelson chain under 
PBC~\cite{hamazaki_19}. Note that a single-chain under OBC does not possess 
such transition because of the real spectrum at weak disorder due to NHSE. 
Here, in the considered two-chain model, the transition occurs under OBC due 
to the interplay of $K$ and $g_{\sigma}$'s. At weak $W$, the $f_{\rm im}$ 
converges to unity as system size increases and with increases in $W$ the 
$f_{\rm im}$ approaches zero signifying the complex-real transition. This 
confirms that the real spectrum at strong $W$ shown in Fig.~1(c) is robust to 
change in $L$. Since the system sizes considered in the present work are small 
and the extrapolation to an infinite system size is difficult, hence we refrain 
from extracting the critical disorder strength $W_c$ using finite-size scaling 
approach. It is noteworthy a similar argument also holds for a Hermitian MBL 
system where an asymmetric scaling is predicted to govern the localization 
transition~\cite{mace_19,garcia_20,laflorencie_20,morningstar_20,abanin_21}. 
Considering the subtleties of finite-size scaling near MBL transition, here we 
tentatively identify the $W_c$ of the localization transition by the crossing 
of $f_{\rm im}$-curves for the largest system sizes available. This lead to 
$W_c\approx 9$ (see the crossing of curves for $L=7$ and $L=8$) at $K=1$ and 
$g_{\uparrow}/g_{\downarrow} = 0.2$, which is consistent to the similar analysis 
done using complex level spacing ratio in the main text [Fig.~\ref{rbar1}(b)]. 

We now discuss the role of boundary conditions in terms of the model parameters 
$K$ and $W$. The disorder-averaged $f_{\rm im}$ for $L=6$ is shown in 
Fig.~\ref{eigen_w}. For larger system sizes, the qualitative behaviour of the 
energies does not change, however the critical value of the transitions might 
vary.
\begin{figure}[ht]
  \includegraphics[width=\linewidth]{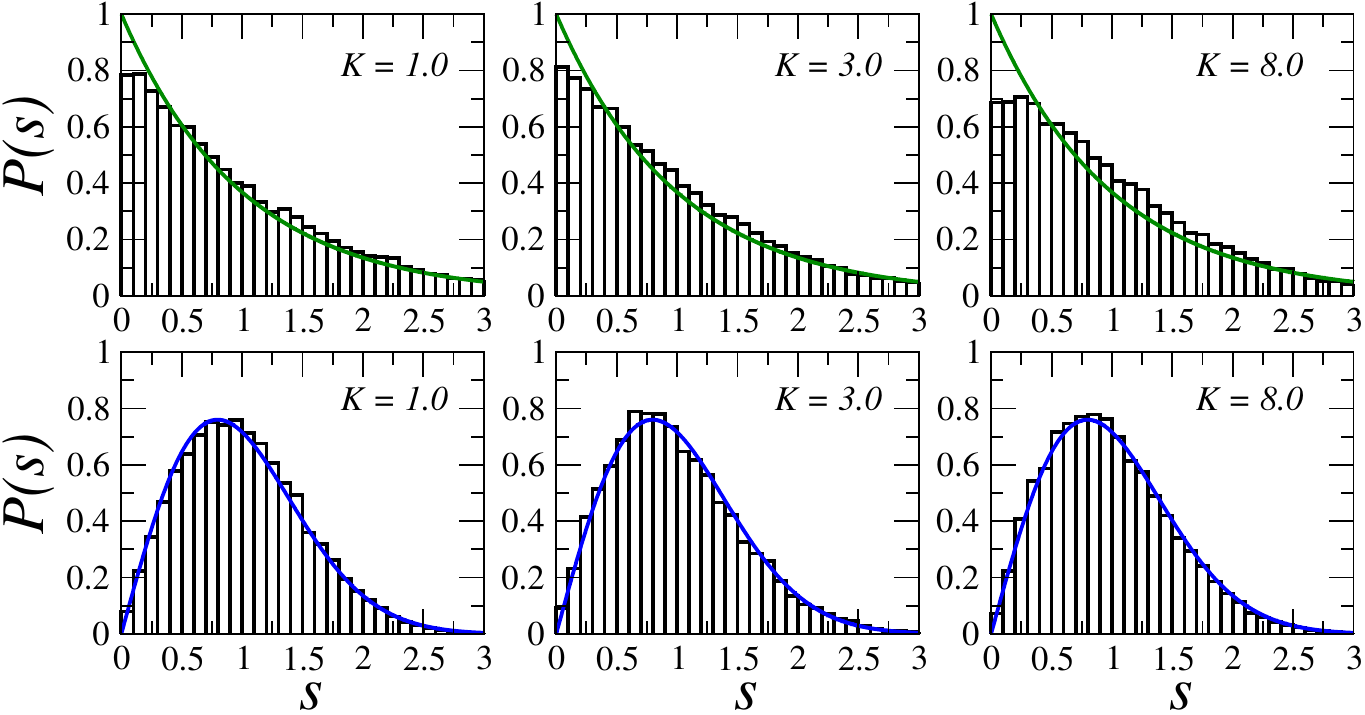}
  \caption{The nearest-neighbour level spacing distributions as a function of
           the inter-chain coupling strength $K$ for
           $g_{\uparrow} = g_{\downarrow}$. The upper (lower) panel represents
           the distributions with correlated (uncorrelated) disorder for the 
	   system size $L=8$. Here, the disorder strength $W=2$ and the 
	   distributions are averaged over $100$ disorder realization. The green
	   and blue lines represent the (real) Poisson and GOE level-spacing 
	   distributions, respectively.}
\label{level_dist}
\end{figure}
\begin{figure}[ht]
  \includegraphics[width=\linewidth]{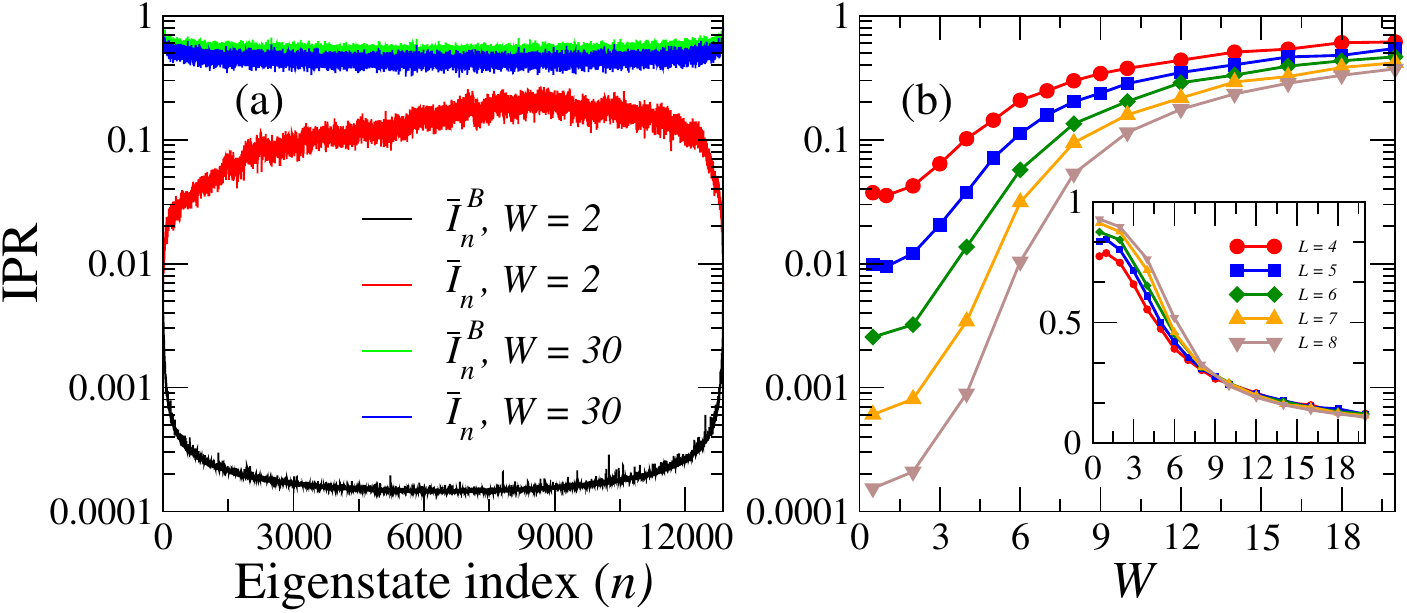}
  \caption{(a) The disorder-averaged inverse participation ratio obtained using
           $n$th right eigenstates $(\bar{I}_n)$ and biorthogonal IPR
           $(\bar{I}_{nB})$ for open coupled chains of $L = 8$. The averaged-IPR
	   is shown for two disorder strengths : one in the delocalized $(W=2)$
           and other in the localized $(W=30)$ regime. The other parameters of 
	   the model are $K=1$ and $g_{\uparrow}/g_{\downarrow} = 0.5$. (b) The
           finite-size effects of biorthogonal IPR (averaged over both disorder
           realizations and Hilbert space) as a function of the disorder 
	   strength ($W$). At weak $W$, the IPR approaches zero in the 
	   delocalized regime and at strong $W$, the IPR of different system 
	   sizes converges to one. In the inset of (b), we show the 
	   corresponding biorthogonal fractal dimension for different $L$.}
\label{ipr_plot}
\end{figure}

Under OBC, as discussed previously~(and in the main text), the complex-real
spectral transition induced by inter-chain coupling or by disorder is evident in
Fig.~\ref{eigen_w}(a,b,c). For open coupled-chain with unequal non-Hermiticity
parameters or for hybridized chains ($g_{\uparrow} = 0, g_{\downarrow} \neq 0$
and vice-versa) the system possesses complex energies. While the system remains 
delocalized at $W=2$, it exhibits the complex-real transition due to the 
interplay of non-Hermiticity and inter-chain coupling. Hence, we predict an 
eigenspectral transition, which does not coincide to the localization 
transition. As the $g_{\uparrow}/g_{\downarrow}$ approaches unity, the fraction
of complex energies decreases and at $g_{\uparrow} = g_{\downarrow}$ the system
exhibits real spectrum and is devoid of dynamical instability. For periodic 
chains, the eigenspectrum remains complex as a function of $K$ at weak disorder, 
as illustrated in Fig.~\ref{eigen_w}(d,e,f). The complex nature of the energies 
with PBC is related to the plane-wave character of the eigen-wavefunction and 
the prevailing role of non-reciprocal tunnelings. At strong disorder, the 
wavefunctions are localized~(as ascertain by the eigenstate properties in the 
main text), and delocalization-localization transition coincides to the 
complex-real transition. In short, the $K$-driven transition is absent in PBC 
case whereas the $W$-driven is present for both boundary cases.

%%%%%%%%%%%%%%%%%%%%%%%%%%%%%%%%%%%%%%%%%%%%%%%%%%%%%%%%%%%%%%%%%%%%%%%%%%%%%%%
%%   Appendix B : Level distributions for $g_{\uparrow} = g_{\downarrow}$   %%%
%%%%%%%%%%%%%%%%%%%%%%%%%%%%%%%%%%%%%%%%%%%%%%%%%%%%%%%%%%%%%%%%%%%%%%%%%%%%%%%
\section{Level distributions for $g_{\uparrow} = g_{\downarrow}$}
\label{app_B}
 For $g_{\uparrow} = g_{\downarrow}$, the IGT maps the non-Hermitian 
coupled-chain system to a Hermitian one~\cite{hatano_21}. In the meanwhile, 
the spectral properties strongly depend on the symmetries of disorder potential. 
The correlated~(spin-independent) disorder preserves SU(2) spin-symmetry and 
leads to a Poisson level-spacing distribution at delocalized regime. For the 
Hermitian disordered systems, it has been shown that a symmetry breaking field 
or diagonalization of a single symmetry block of the Hamiltonian leads to 
GOE-like behaviour at weak disorder strength~\cite{mondaini_15,suthar_20}. The 
uncorrelated~(spin-dependent) potential breaks the spin symmetries and results 
into a GOE distribution. The nearest-neighbour level-spacing distributions for 
$W=2$ and $g_{\uparrow}=g_{\downarrow}$ using two different disorder potentials 
are illustrated in Fig.~\ref{level_dist}. 

%%%%%%%%%%%%%%%%%%%%%%%%%%%%%%%%%%%%%%%%%%%%%%%%%%%%%%%%%%%%%%%%%%%%%%%%%%%%%%%
%%                     Appendix C : Inverse participation ratio             %%%
%%%%%%%%%%%%%%%%%%%%%%%%%%%%%%%%%%%%%%%%%%%%%%%%%%%%%%%%%%%%%%%%%%%%%%%%%%%%%%%
\section{Finite-size effects of inverse participation ratio}
\label{app_C}
Here we discuss the disorder-averaged inverse participation ratio as a function
of the eigenstate index. As discussed in the main text, for the non-Hermitian 
systems, the IPR can be defined in two ways: $\bar{I}_{n}$ is defined using 
$n$th left or right eigenstate while $\bar{I}_{nB}$ is using the biorthogonal 
basis. Here the overline denotes the disorder average. Fig.~\ref{ipr_plot}(a) 
shows $\bar{I}_{nB}$ and $\bar{I}_{n}$ in the whole energy spectrum. The 
parameters considered are $K=1$, $g_{\uparrow}/g_{\downarrow} = 0.5$, and 
$L=8$. At weak disorder~$(W=2)$, IPR is small, suggesting delocalization. 
Interestingly, the IPR $\bar{I}_{n}$ shows a finite value while 
$\bar{I}_{nB} \approx 0$ for $W=2$. As explained in the main text, the finite 
$\bar{I}_{n}$ at weak disorder can be attributed to the presence of NHSE. At 
strong $W$, the large values of IPR suggest many-body localization. Furthermore, 
both definitions of IPR give similar values, indicating the suppression of NHSE 
at strong disorder. It is important to note that the choice of eigenstates for 
IPR is only important for open chains. Under PBC, the results from both the 
definitions coincide due to the absence of NHSE. We next present the finite-size 
effects on biorthogonal IPR and corresponding fractal dimension 
in Fig.~\ref{ipr_plot}(b). Here we consider the biorthogonal definitions so that 
the interference of NHSE-induced phenomena can be eliminated and we could 
inspect system-size effects on the role of disorder potential. At weak disorder 
strengths, the averaged-IPR decreases with increase in $L$. The contrast 
behaviour is also seen for biorthogonal fractal dimension. Hence, 
$I^{B}_{\rm avg} \rightarrow 0$ (or $\eta_{B} \rightarrow 1$) with increase in 
$L$ exhibits the delocalized regime. As $W$ increases, the IPR tends to 
approach unity, and converse behaviour of FD is evident from the inset in 
Fig.~\ref{ipr_plot}(b). The convergence in IPR and FD for different system 
sizes confirms the disorder-driven many-body localization of non-Hermitian 
coupled chains.

\bibliography{nh_mbl}{}
\bibliographystyle{apsrev4-1}
\end{document}